\documentclass[twocolumn]{aastex62}

\graphicspath{{./}{figures/}}

\received{November 7, 2018}
\revised{January 25, 2019}
\accepted{February 16, 2019 }

\shorttitle{Cosmic transient detection}
\shortauthors{Varghese et al.}

\begin{document}

\title{ DETECTION OF A LOW FREQUENCY COSMIC RADIO TRANSIENT USING TWO LWA STATIONS}

\correspondingauthor{Savin Shynu Varghese}
\email{savin@unm.edu}

\author{S. S. Varghese}
\affil{University of New Mexico, Albuquerque, NM, USA}

\author{K. S. Obenberger}
\affiliation{Air Force Research Laboratory, Kirtland AFB, NM, USA}

\author{J. Dowell}
\affiliation{University of New Mexico, Albuquerque, NM, USA}

\author{G. B. Taylor }
\affiliation{University of New Mexico, Albuquerque, NM, USA}





\begin{abstract}
We report the detection of a potential cosmic radio transient source using the two stations of the Long Wavelength Array. The transient was detected on 18 October 2017 08:47 UTC near the celestial equator while reducing 10,240 hours of archival all-sky images from the LWA1 and LWA-SV stations. The detected transient at 34 MHz has a duration of 15 - 20 seconds and a flux density of 842 $\pm$ 116  Jy at LWA1 and 830 $\pm$ 92 Jy at LWA-SV. The transient source has not repeated, and its nature is not well understood. The Pan-STARRS optical telescope has detected a supernova that occurred on the edge of the position error circle of the transient on the same day. 
\end{abstract}



\section{Introduction} \label{sec:intro} 

Radio transient sources can be defined as a class of objects which emit radio waves in the form of bursts, flares or pulses from short duration (less than a few seconds) to long durations (greater than a few seconds). The progenitors of such sources are usually associated with explosive or dynamic events. Probing such sources helps to understand the physical mechanisms of these extreme energetic events \citep{cord04}. We can classify transients as extragalactic, galactic and atmospheric based on the location of their occurrence.

 Most transients have been discovered through high time resolution (less than a second) observations and blind imaging of the sky. The high time resolution studies at high frequencies have discovered giant pulses from the Crab pulsar  at 5.5 and 8.6 GHz \citep{2003Natur.422..141H}, single dispersed bursts from rotating radio transients (RRAT; \citealt{http://adsabs.harvard.edu/abs/2006Natur.439..817M}) at 1.4 GHz and the new class of Fast Radio Bursts at 1.4 GHz \citep{http://adsabs.harvard.edu/abs/2007Sci...318..777L}. Thirteen new FRBs  have been detected between 400 -800 MHz by the Canadian Hydrogen Intensity Mapping Experiment (CHIME) Collaboration \citep{https:// doi.org/10.1038/s41586-018-0867-7, https://doi.org/10.1038/s41586- 018-0864-x}. Several high time resolution observation campaigns have been conducted at low frequencies below 300 MHz searching for giant pulses from pulsars, RRATs and FRBs. At low frequencies, giant pulses from pulsars have been detected, but the detection rate is low for RRATs and zero for FRBs \citep{http://dx.doi.org/10.3847/0004-637X/829/2/62,http://dx.doi.org/10.3847/0004-637X/831/2/140,  http://dx.doi.org/10.1088/0004-637X/809/1/67, doi:10.1093/mnras/stv1306,http://dx.doi.org/10.1051/0004-6361/201424495}. The scattering of the radio pulses due to inhomogeneities in the medium can cause temporal smearing of the pulse to longer durations at low frequencies. This may limit the detection of short duration transients in the high time resolution observations. This makes fast imaging of the sky on timescales of few seconds a good option for capturing scatter broadened pulses at low frequencies \citep{http://dx.doi.org/10.1088/2041-8205/776/1/L16,10.1093/mnras/stt1598,10.1093/mnras/stw451}.

In the past few decades, blind searches of the sky focused at frequencies above 300 MHz have discovered galactic center transients, bursts from ultra cool dwarfs and flare stars, day scale transient in Spitzer-Space-Telescope Wide-area Infrared Extragalactic Survey (SWIRE) Deep Field: 1046+59 and 15 transients in the Molonglo Observatory Synthesis Telescope (MOST) transient survey \citep{http://adsabs.harvard.edu/abs/2005Natur.434...50H,http://adsabs.harvard.edu/abs/2007ApJ...663L..25H,jac,jag,ban}. The transient radio sky below 300 MHz is not well studied  and remains poorly explored below 100 MHz.  Fast imaging techniques on shorter time scales are required to capture transient pulses at low frequencies. The initial study of transients were limited by the narrow field of view (FoV) of the radio instruments. With advances in technology, however, new low frequency radio instruments have a wide field of view, increased bandwidth, sensitivity to study the dynamic transient sky. The currently operating major low frequency radio telescopes include the International Low-Frequency Array (LOFAR; \citealt{van}), the Murchinson Wide Field Array (MWA; \citealt{http://dx.doi.org/10.1017/pasa.2012.007}) and the Long Wavelength Array (LWA; \citealt{2012JAI.....150004T,http://arxiv.org/abs/1307.0697}).

Several sources have been theorized to emit radio pulses but are yet to be detected. This includes low frequency prompt emission from GRBs  \citep{us,sag}, exoplanets \citep{far}, giant flares from magnetars or extragalactic pulsars \citep{http://adsabs.harvard.edu/abs/2003ApJ...596..982M} and annihilating black holes \citep{rees}. Recently, several observing campaigns have been carried out to image the transient sky at low frequencies on integration time scales from 5 s to several hours.

\citet{http://adsabs.harvard.edu/abs/2016MNRAS.459.3161C}  conducted a transient search from 115-190 MHz using LOFAR with cadences between 15 min to several months. No significant transient was found after analyzing 151 images with  sensitivity greater than 0.5 Jy obtained from  2275 deg square survey area.  

\citet{2016MNRAS.456.2321S} detected a new low frequency radio transient at 60 MHz after 400 hrs of monitoring of the North Celestial Pole (NCP) in the LOFAR Multi Snapshot Sky Survey (MSSS).  The identified transient had a flux density of 15-25 Jy with a duration of few minutes. The transient was not found to repeat after follow-up observations and did not have any obvious optical or high energy counterparts. 

\citet{http://dx.doi.org/10.1093/mnras/stt2200} carried out a transient search on characteristic time scales of 26 min and 1 year with MWA at 154 MHz covering 1430 square degree FoV. The search did not identify any transient sources greater than 5.5 Jy in 51 images obtained from six days of observations.

 \citet{10.1093/mnras/stw451} searched for transient and variable sources using MWA at 182 MHz. No transients were detected on time scales from 28 s to 1 year with flux density greater than  0.285 Jy.

\citet{10.1093/mnras/stw3087} conducted a transient search on timescales from 1 to 3 year by comparing the 147.5 MHz TIFR GMRT Sky Survey Alternative Data Release 1 (TGSS ADR1) and the 200 MHz GaLactic and Extragalactic All-sky Murchison Widefield Array (GLEAM) survey catalogs.  The search found a transient source with a flux denisty of 182 $\pm$ 26 mJy in the TGSS ADR1 which was not present in the GLEAM survey.


Using the first station of LWA,  \citet{2014ApJ...785...27O} detected two kilojansky flux density transient sources while searching for low frequency prompt emission from gamma ray bursts. These sources were detected at 37.9 MHz and 29.9 MHz with a duration of a few minutes. The transient search was carried out using the all-sky imaging capabilities of the LWA All-Sky Imager (LASI; \citealt{2015JAI.....450004O}). Follow-up observations with optical cameras revealed that the radio emission is temporally and spatially associated with optical meteors \citep{2014ApJ...788L..26O}. These meteor radio afterglows (MRA) begin to emit within a few seconds after the optical activity and they can be classified as a new form of atmospheric transient.
MRAs were studied extensively to understand the origin and energetics of the emission. The current understanding is that these broadband, non-thermal radio sources are the result of electromagnetic conversion of electrostatic plasma waves within the turbulent plasma of meteor trails \citep{2015JA021229}.

With a detection rate of 60 MRAs per year, it is difficult to differentiate these foreground sources with events of cosmic origin using a single LWA station. The earlier transient studies using a single LWA station \citep{2014ApJ...785...27O, 2014ApJ...788L..26O} assumed that all unpolarized transients lasting from few seconds to few minutes duration as MRAs. However, some of the events assumed to MRAs, might have been cosmic in nature but there was no way to properly identify the transients not directly associated with an optical meteor. The recent commissioning of the new LWA station at Sevilleta National Wildlife Refuge (LWA-SV) provides a new opportunity to observe cosmic transients. The two stations are separated by 75 km which is sufficient to differentiate the foreground transient events like lightning, MRAs, radio frequency interference (RFI) and low earth orbit satellites from cosmic events, while still being close enough to share over 99\% of the sky. So far LOFAR MSSS is the only low frequency survey that has carried out the transient search close to the LASI operating frequency with wide FoV.

In this paper, we present a two year study of all-sky images from both LWA stations which has identified a new promising cosmic transient candidate. Sections 2 and 3 describe the observations and data reduction methodology. Section 4 describes the detection of cosmic transient candidate event. Section 5 gives an extensive analysis of the common transient events observed in both LWA stations and explains why one transient event is a statistically significant and a promising candidate.






\newpage
\section{Observation} \label{sec:obs}
The first station of the Long Wavelength Array (LWA1) is a low frequency radio telescope located in central New Mexico \citep{{2012JAI.....150004T}}. The telescope operates between 10 and 88 MHz frequency range and it is collocated with the Karl G. Jansky Very Large Array (VLA). The array is comprised of 256 dual polarization dipole antennas along with five additional outrigger antennas located at 200-500 m distance from the center of the array. The core of the array is distributed in the form of a 100 $\times$ 110 m ellipse. 

The second station, LWA Sevilleta (LWA-SV) was commissioned in November 2017 \citep{{2017JAI.....650007C}}. LWA-SV is located at the Sevilleta National Wildlife Refuge, 75 km North East of LWA1. LWA-SV has a similar layout to LWA1 and the backend hardware has similar but not identical capabilities. 

Both the stations primarily operate in two modes, digital beamforming and the all-sky mode. In the digital beamforming mode, a time domain delay-and-sum architecture is used to form beams. The delay processed signals from each antenna can be added to form up to 4 independently steerable dual polarization beams at any direction in the sky. Each beam can be tuned to two central frequencies within the operating range of the telescope with a bandwidth up to 19.6 MHz in LWA1 and 9.8 MHz in LWA-SV. 

The all-sky mode takes advantage of the primary beam of a single dipole antenna which is sensitive to the whole sky. The all-sky monitoring is done in Transient Buffer Wide (TBW) and Transient Buffer Narrow (TBN) mode. In the TBW mode, the voltage time series from each antenna is collected at the entire 78 MHz bandwidth for 61 ms and it takes 5 minutes to write out the data. TBN mode collects the voltage series time series data from each antenna continuously at 100 kHz bandwidth and can be tuned to anywhere in the operating frequency of the stations. The collected data is then sent to a software FX correlator \citep{http://arxiv.org/abs/1307.0697}.

LWA All-Sky Imager (LASI) is the back end correlator for the both LWA stations \citep{{2015JAI.....450004O}}. LASI cross correlates real time TBN data from each antenna and produces an all-sky image every 5 seconds. The produced images are uploaded to LWA TV website\footnote{\url{http://www.phys.unm.edu/~lwa/lwatv.html}} and stored in the LWA archive\footnote{\url{https://lda10g.alliance.unm.edu/}}. For this work, we have used over 10,240 hours (May  2016 - July 2018) of data recorded from each LWA station at 34 MHz and at 38 MHz.







\section{Data Reduction} \label{sec:data}

\subsection{Transient Pipeline}
The transient search pipeline uses an image subtraction algorithm to find the transient candidate events from both stations \citep{{2015JAI.....450004O}}. In the image subtraction process, an average of the previous 6 images is subtracted from the running image. At the same time the script masks out the bright radio sources like Cyg A, Cas A for efficiently finding transients. The pixels with flux density greater than 6 $\sigma$ in the subtracted image are marked as transient candidates. The detection threshold varies near the Galactic plane and has been discussed in \citep{{2015JAI.....450004O}}. Currently the transient search is carried out on 5, 15 and 60 s integrations in Stokes I and V. 
\subsection{Comparison of Transient events}
The pipeline outputs the time and coordinates of the transient events detected from each LWA station.  The noise of the LWA1 subtracted image measured at  38 MHz is 41 Jy at zenith and increases towards the horizon \citep{2015JAI.....450004O}. The LWA-SV station is 5-10\% more sensitive than the LWA1 station because it has more fully functioning dipoles.  Depending on the location of a transient event occurring from the zenith of each station, image noise changes and leads to some time difference in detecting them at each station. We compare the output files from both stations to find the associated events which occur within 30 seconds difference.

The next step is to look for cosmic transient candidates and meteor afterglow candidates. If the angular difference between coordinates of events detected from each station is less than 3 degrees, then it is classified as a cosmic transient candidate. Since cosmic transient events occur at great distances compared to the 75 km baseline, the angular direction to the event from each station would be the same. If the angular difference is greater than 3 degrees, then it is classified as a MRA candidate. The 3 degree angular difference threshold is given in order to account for the pointing of telescope and random errors from ionospheric disturbances. For a 75 km baseline, 3 degree angular difference corresponds to a distance of 1400 km.

The main advantage of this method is to detect common events which can be cosmic or meteor afterglow candidates. Also at the same time it removes nearly all the local RFI effects arising from power lines, lightning, air planes, etc. However, it still identifies some false positive events like scintillation and radio transmitter signal reflections from meteor trails.

The whole process of finding transients using the two stations is automated. Once LASI collects the all-sky images for a day, the transient search pipeline processes all the data and finds the transient candidates. At the end of each UT day, the collected events from each station are compared, and events that are classified as either a MRA or a cosmic transient are emailed to the authors of this study.


\section{cosmic transient candidate Detection} \label{sec:det}

The radio transient candidate LWAT 171018 was detected after analyzing the archival  all-sky images from the two LWA stations. The events took place on 18 October 2017 (MJD 58044) 8:47:33 UTC in LWA-SV and 8:47:38 in LWA1. The LASI correlator was collecting the all-sky images at 34 MHz in both stations. Each station recorded the event in the adjacent time bins where each bin is a 5 second integration. The event detection in each station can be considered to be simultaneous within the uncertainty of our measurement.  The top panel in Fig. 1 shows the Stokes I light curves of the transient event seen from each station. 

\begin{figure*}
    \noindent \hspace{4cm}
     \hspace{7.5cm}
\gridline{\fig{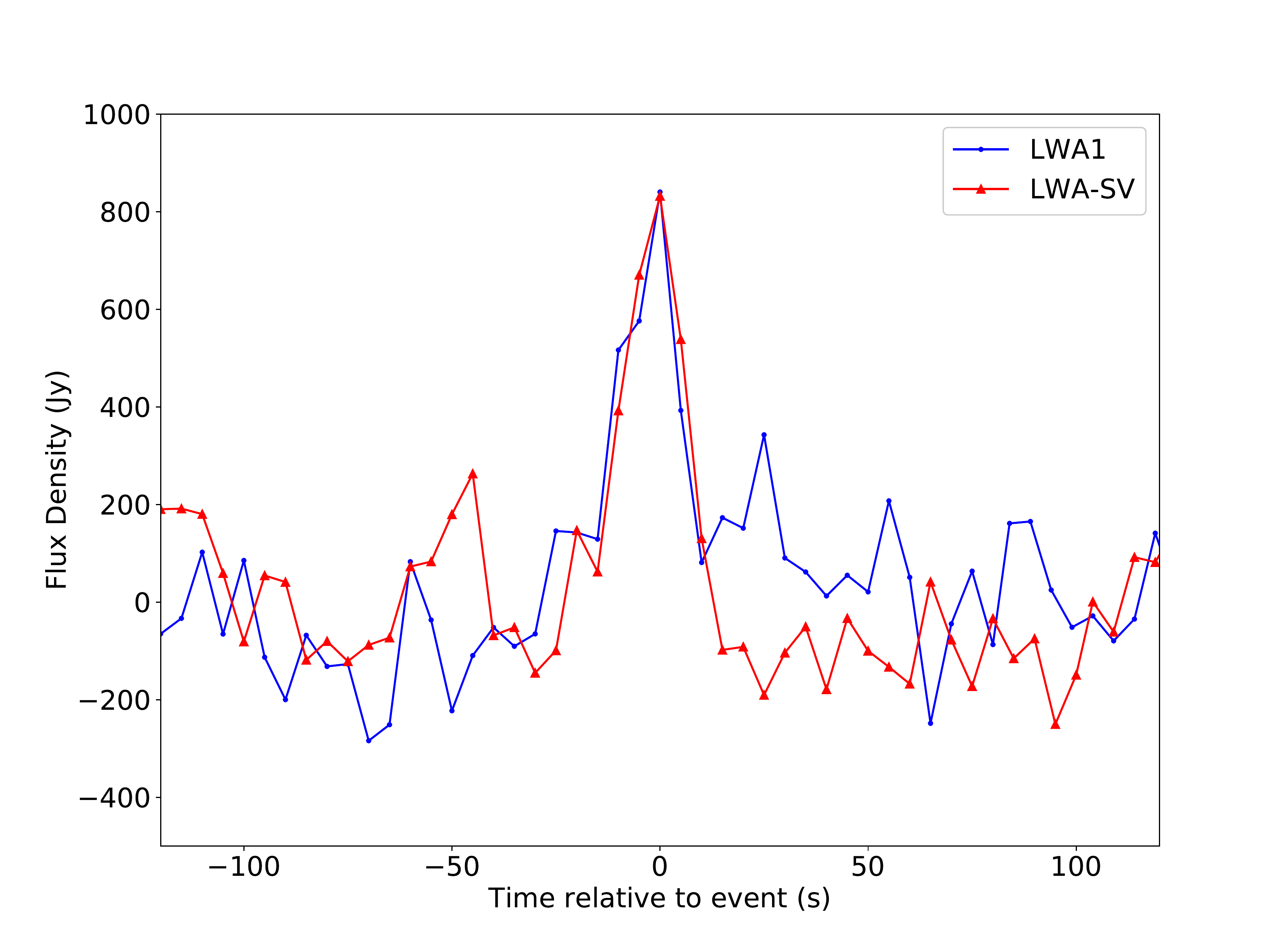}{0.7\textwidth}{}}
\gridline{\fig{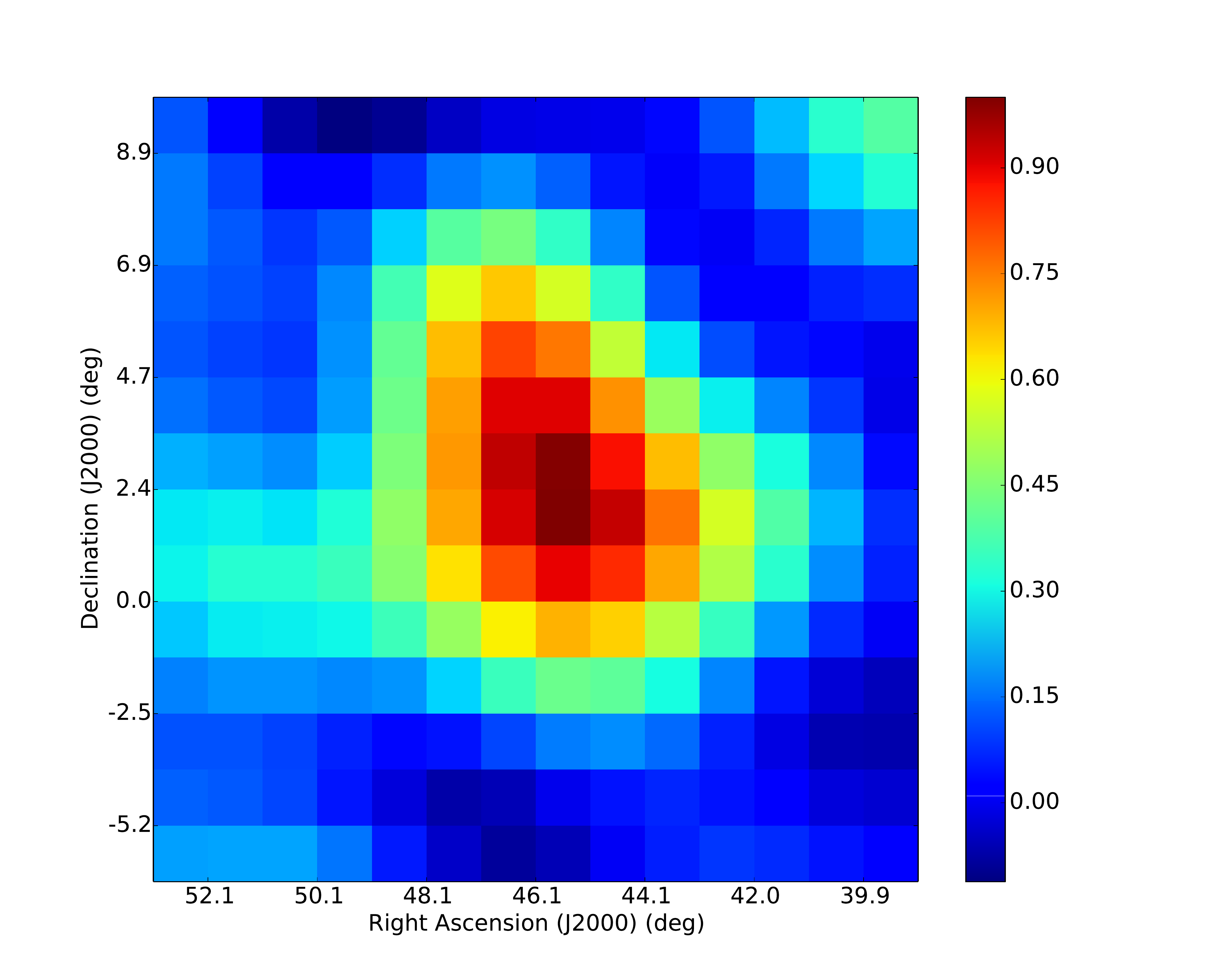}{0.45\textwidth}{LWA1}
          \fig{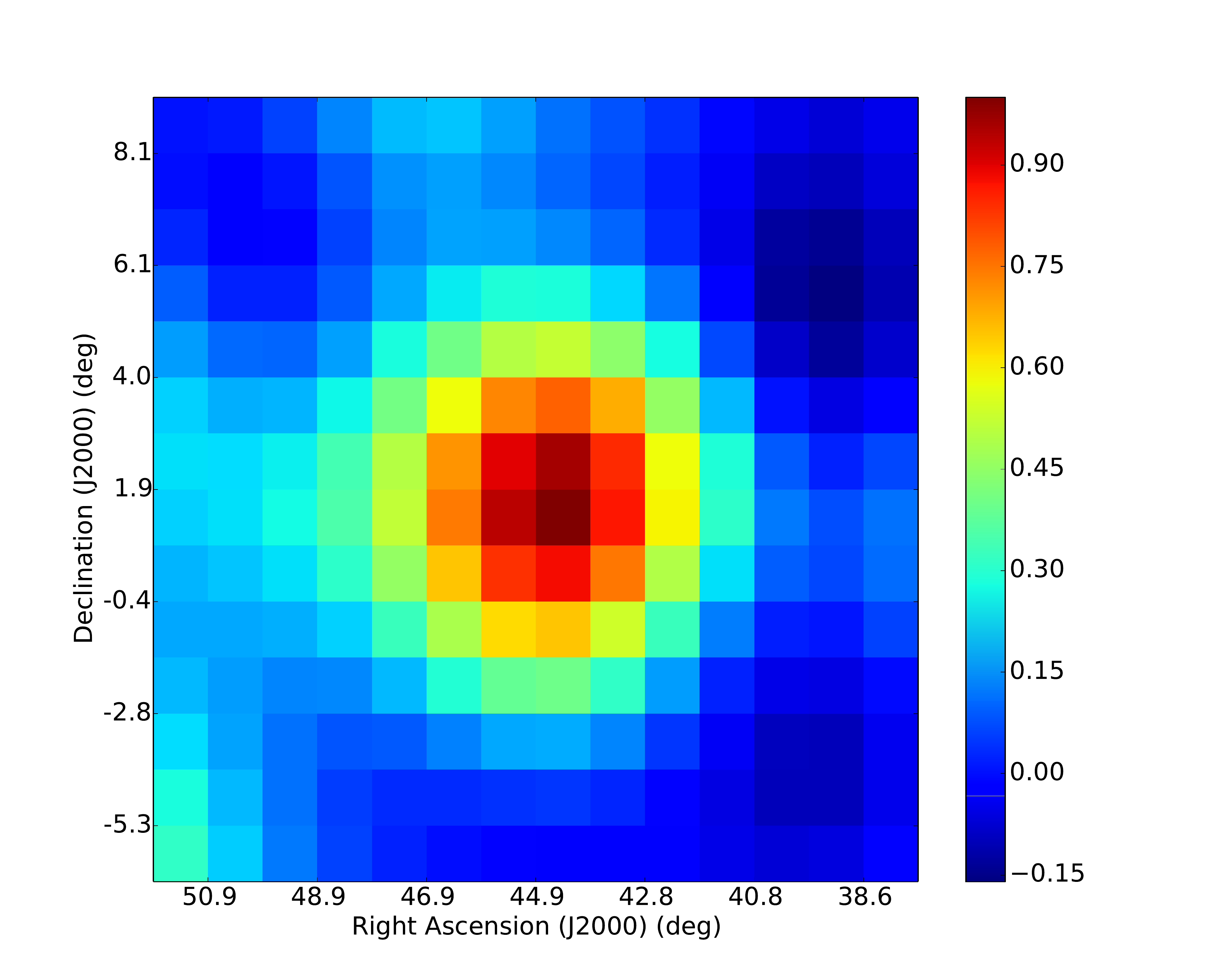}{0.45\textwidth}{LWA-SV}
          }

\caption{The Stokes I light curves of the transient event LWAT 171018. The blue light curve denotes LWA1 and the red curve denotes LWA-SV.  The bottom panel shows the subtracted image of the transient from each station. The color bar shows the normalized pixel values in the subtracted image. Each pixel in the image corresponds to 1.016 degrees on the sky.}
\end{figure*}

There is difference in the signal to noise in both stations due to the difference in the number of working antennas. The light curves shows that the emission lasted for 15 - 20 seconds in each station. LWA1 has recorded 7.24 $\sigma$ source signal and LWA-SV has 8.81 $\sigma$ detection from the all-sky image indicating that the emission is relatively faint. The bottom panel in Fig. 1 shows the subtracted image of the transient seen from LWA1 and LWA-SV which suggest that it is a point source. There is more noise in the LWA1 image compared to LWA-SV. The different ionosphere above each station and the noise being added during averaging in the image subtraction may lead to small difference in apparent source structure which is evident from the images. 

The all-sky image from the time of peak emission was used to accurately measure the flux density. The average of 10 noise-like images is subtracted from the peak flux image to measure the peak flux density of transient and thermal noise in arbitrary units. The thermal noise is calculated by the standard deviation from a quiet portion of the subtracted image. The flux and noise values were calibrated using the bright radio source Cyg A,  scaling them to Jansky. The measured value of transient flux density from the LWA1 is 842 $\pm $ 116 Jy and at LWA-SV is 830 $\pm $ 92 Jy. The calculated error bars are thermal noise values from the peak flux image.  

\section{How to confirm the Transient?}
The  presence of similar light curve patterns and close flux density values is not sufficient evidence by itself to confirm a cosmic origin. In the automated transient search pipeline, the comparison script looks for power spikes happening in both stations which are within a 5 s interval. The power spike at the same time in both stations could have a number of origins. Below we examine each of the possible origins.
\paragraph{Meteor Radio Afterglows}
The MRA events usually occur at 90 - 130 km elevation. The difference in angular direction to the event from each station can vary from 30 to 45 degrees in the sky as the two stations are separated by 75 km. Therefore the two station will not record MRAs in same angular directions (Right ascension, Declination) and they can be ruled out.

\paragraph{Radio frequency interference}
These are mostly man made signals reflecting off the ionosphere and meteor plasma trails. The origin of RFI seen in both stations can be from the same or different transmitters. The reflection events are typically bright, short in duration, highly linearly or circularly polarized and are narrow band in frequency. Fig. 2 shows the light curves of the event at stokes Q, U and V from both stations. The all-sky image data is collected at 100 kHz bandwidth and the spectrum information is not available as the measurement sets are deleted after one month from the day of observation. This limits looking into the raw data for narrow band RFI events. But the lack of a polarized detection in both stations suggests that we can rule out the case of coincident RFI.


\begin{figure*}
    \noindent \hspace{4cm}\text{LWA1}
     \hspace{7.5cm}\text{LWA-SV}

\gridline{\fig{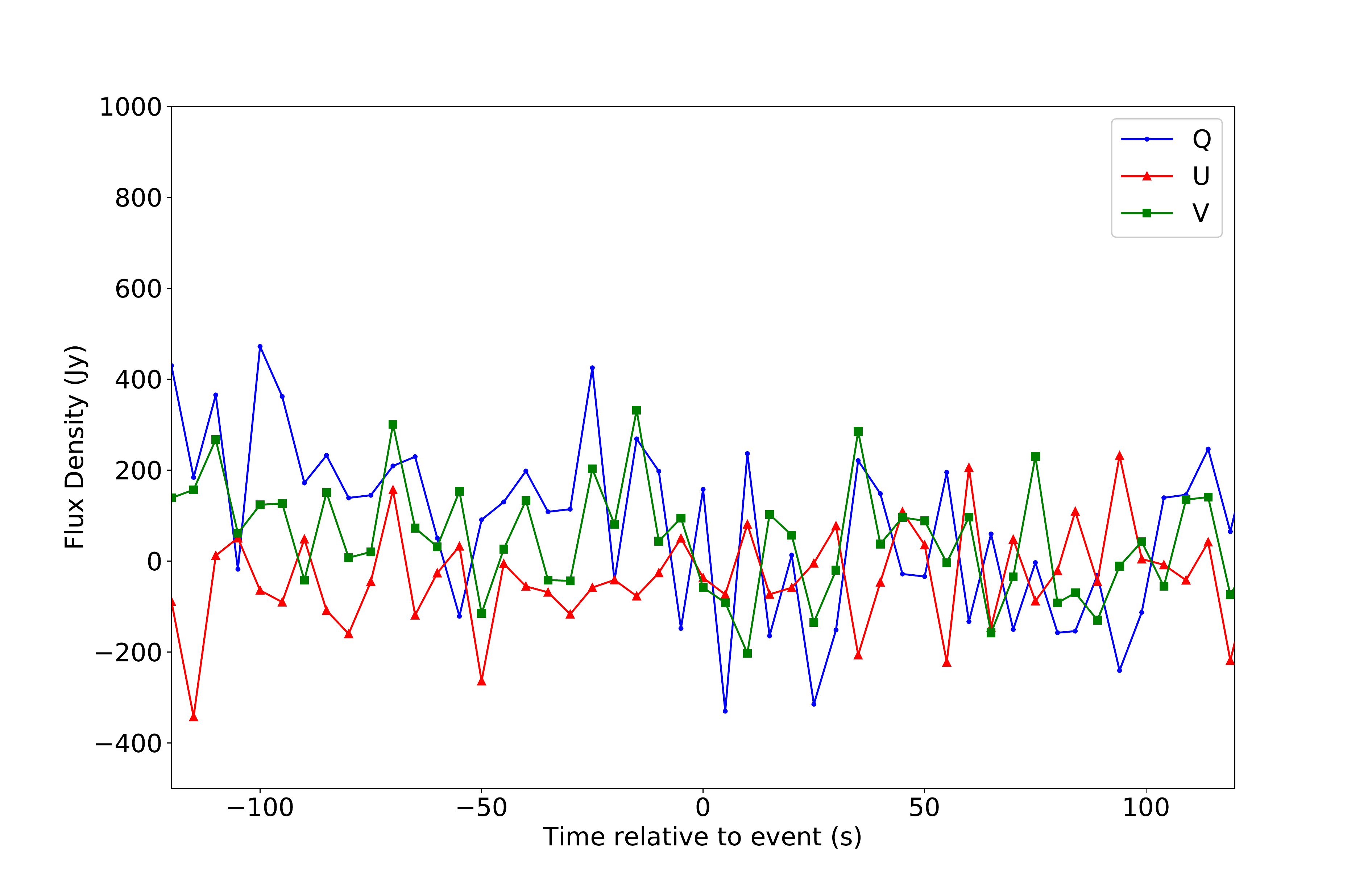}{0.45\textwidth}{}
          \fig{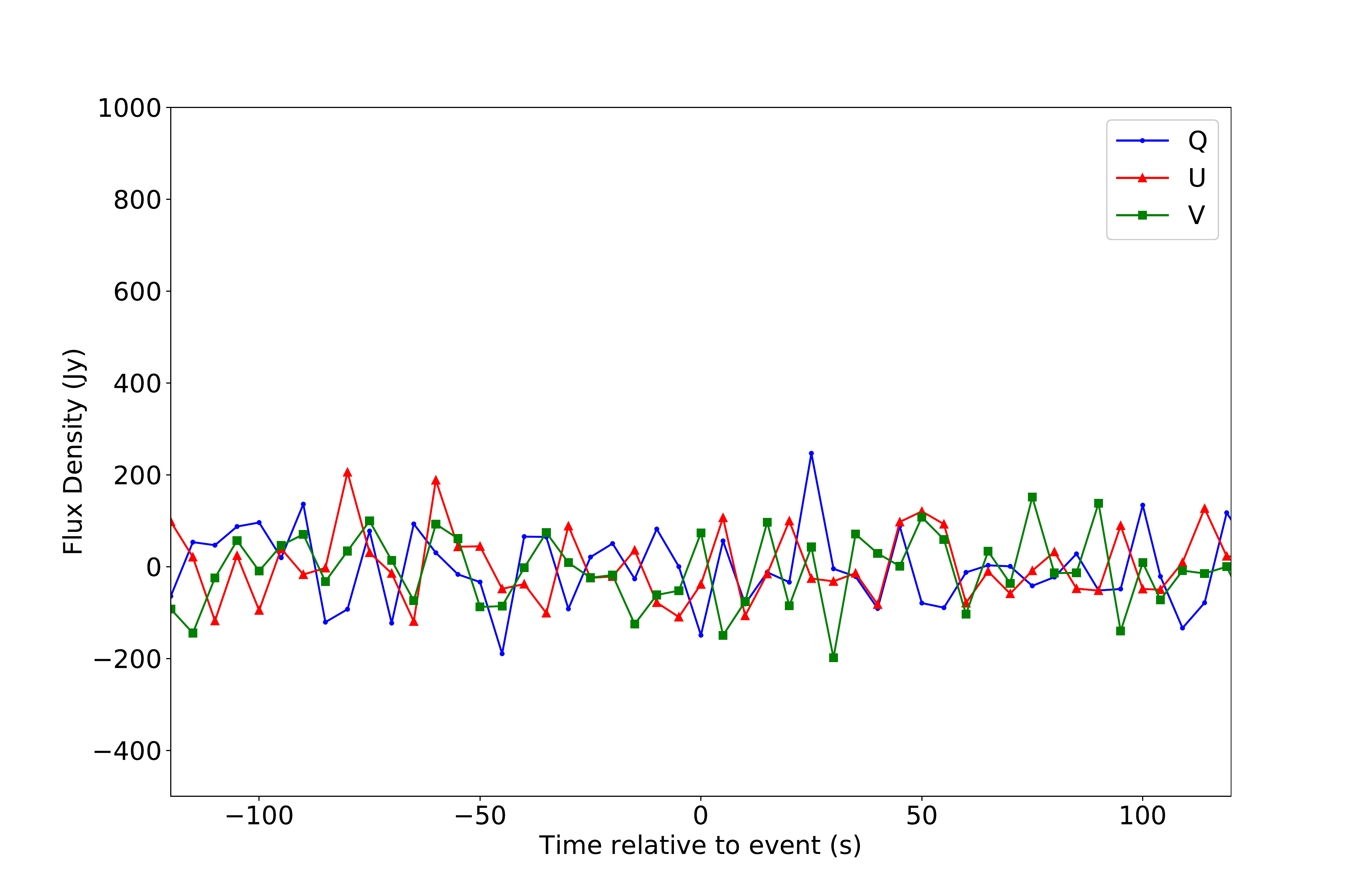}{0.45\textwidth}{}
          }
  
\caption{The Stokes Q, U and V light curve of the transient event from LWA1 (left) and LWA-SV (right).}
\end{figure*}

\paragraph{Scintillation}

Scintillation of bright radio sources by Earth's ionosphere is a problem at lower frequencies \citep{{2015JAI.....450004O}}. The ionosphere contains magnetized plasma and density variations, which cause rapid changes in observed flux (up to a factor of 15) and can offset the position of sources by few degrees. This effect becomes intense for bright compact sources and at the same time sources below the nominal detectable limit can appear above the noise floor for some period of time. The scintillation seen in each station can be due to same or different radio sources. In order to reduce false transient events due to scintillation, the script masks radio sources brighter than 50 Jy from the VLA Low Frequency Sky Survey at 74 MHz (VLSS; \citealt{2007AJ....134.1245C}; \citealt{2012RaSc...47.0K04L}). This removes a significant portion of the sky ($\approx $12\%) but is the best way to avoid the confusion between transients and scintillation.

A full statistical analysis determining the rates of scintillation based on sky position, flux density and source structure are beyond the scope of this paper. However, anecdotal evidence suggests that sources as low as 10 Jy (at 74 MHz) can scintillate to detectable levels. It is therefore helpful to calculate the probability that a random transient will be spatially coincident with a VLSS source with flux density greater than 10 Jy at the same LST of LWAT 171018 detection. Using a Monte Carlo simulation with $10^{5}$ beams and the VLSS catalog, we estimate a 15$\%$ chance that a VLSS source $>$ 10 Jy will be within the position error of a random transient.

Typical scintillation light curves are characterized by  random fluctuations with several peaks appearing over a period of about 30 minutes to a few hours. The transient search algorithm may identify these peaks as transients. While scintillating sources often trigger a single station transient, a single source typically does not experience a scintillation spike at both stations at the same time. However, during periods with exceptionally high scintillation double station triggers can occur, these triggers then show up as potential cosmic transients. In the data presented in this paper, we have observed 18 cases of double station coincident source scintillation. Such cases are easy to identify due to their characteristic light curves, and the fact that they typically occur during periods of high scintillation, where many other sources are scintillating at the same time. We have also identified one case of coincident RFI event in both stations.

A statistical approach was required to study the nature of scintillation events and to differentiate them from a real cosmic transient events. For this study, we chose two cases based on their occurrence at the same time and high flux density levels. The first case is our promising, transient event LWAT 171018. The second case is the scintillation candidate MJD 58040. The details of all the scintillation and RFI events are given in the Table 1.  The nature of the unknown event in LWA1 from MJD 58238 is not clear. This could be an MRA event seen by LWA1 which was not in the shared sky region of LWA-SV.

\startlongtable
\begin{deluxetable*}{c|ccccc}
\tablecaption{List of cosmic transient candidate events detected from both LWA stations and their classification \label{tab:table}}
\tablehead{
\colhead{MJD} & \colhead{UTC Time} & \colhead{LWA1} & \colhead{LWA-SV}
& \colhead{Kurtosis LWA1} & \colhead{Kurtosis LWA-SV}}
\colnumbers
\startdata
58019 & 01:42 & RFI & RFI & 1.153 & 37.438\\
58039 & 05:43 & Scintillation & Scintillation & 0.719 &  4.324\\
58040 & 05:26 & Scintillation candidate & Scintillation candidate & 2.067 & 4.817\\
58044 & 08:47 & LWAT 171018    & LWAT 171018 & 0.056 & 0.161\\
58054 & 05:36 & Scintillation & Scintillation & 1.447 & 5.386 \\
58064 & 15:28 & Scintillation &  Scintillation & 132.226 &  17.766\\
58066 & 04:30 & Scintillation &  Scintillation & 5.133 & 1.834\\
58067 & 12:33 & Scintillation &  Scintillation & -0.0445 & 1.612\\
58094 & 08:06 & Scintillation &  Scintillation & 0.664 & 2.201\\
58102 & 05:32 & Scintillation &  Scintillation & 2.537 & 73.111\\
58102 & 11:15 & Scintillation &  Scintillation &  2.534 & 27.901\\
58113 & 06:42 & Scintillation &  Scintillation & 0.639 & 5.354 \\
58128 & 09:35 & Scintillation &  Scintillation & 1.033 & 1.966 \\
58174 & 08:31 & Scintillation &  Scintillation & 4.282 & 1.191\\
58238 & 02:58 & Unknown &  Scintillation & 0.538 & 73.279 \\
58238 & 04:57 & Scintillation & Scintillation  & 2.167 &  6.674\\
58238 & 05:01 & Scintillation & Scintillation & 4.779 & 3.721\\
58341 & 10:25 & Scintillation & Scintillation & 0.672 & 2.177\\
58356 & 17:14 & Scintillation & Scintillation & 0.675 & 1.449\\
\enddata

\end{deluxetable*}
Four methods are used here to analyze the scintillation candidate MJD 58040 and LWAT 171018 to understand their significance. 

The first method is to look at the light curves of the events from each station as well as the averaged light curves. For a light curve with a Gaussian noise, averaging of the light curves from both stations will increase the signal to noise ratio for a real signal.

Fig. 3 shows the light curves from each station and their average for LWAT 171018. In the light curves, the event is defined as the time from 10 s before and after the peak flux point which is denoted as zero second. Noise is defined as all the points in light curve other than the event. The light curves from each station has Gaussian noise and similar peak flux density at the same time. The SNR ratio has increased significantly in the averaged plot curves. Fig. 4 shows the light curves from the scintillation candidate MJD 58040. In the light curves plots, the noise is fluctuating with random peaks over the course of more than an hour. Adding the light curves from both stations has increased the SNR ratio for the scintillation candidate MJD 58040. Even though the SNR ratio has increased, the light curves still has random fluctuations as high as the peak signal.

\begin{figure*}
\epsscale{1.25}
\plotone{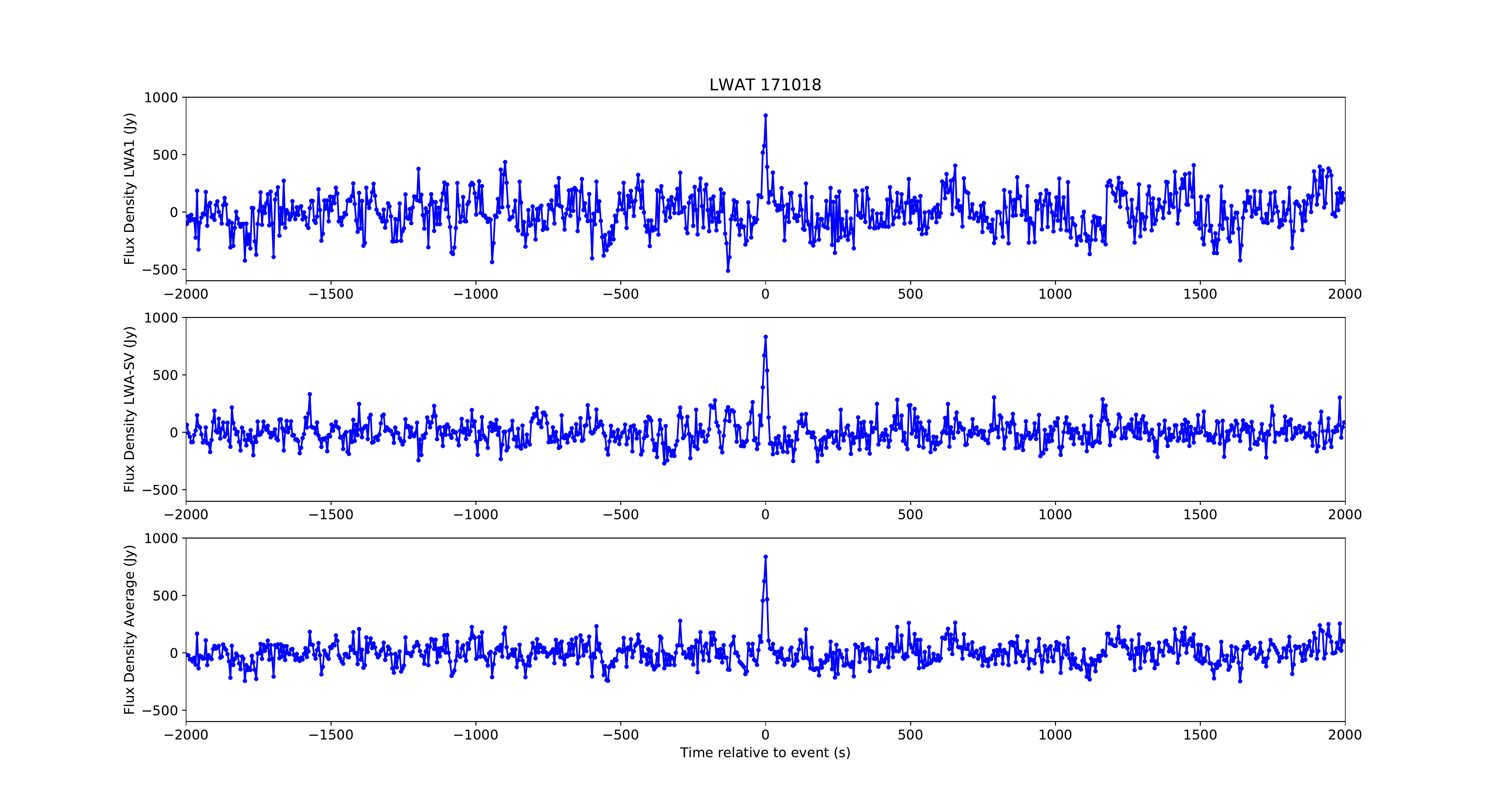}


\caption{Plot showing the light curves of LWAT 171018 on a longer time axis. The first top panel shows the light curve of event from LWA1 with SNR = 5.28. The middle panel shows the light curve from LWA-SV with SNR = 8.44. The bottom panel shows the average light curve from both stations with an improved SNR = 9.18. The time zero denotes the peak time of the event.}
\end{figure*}

\begin{figure*}
\epsscale{1.25}
\plotone{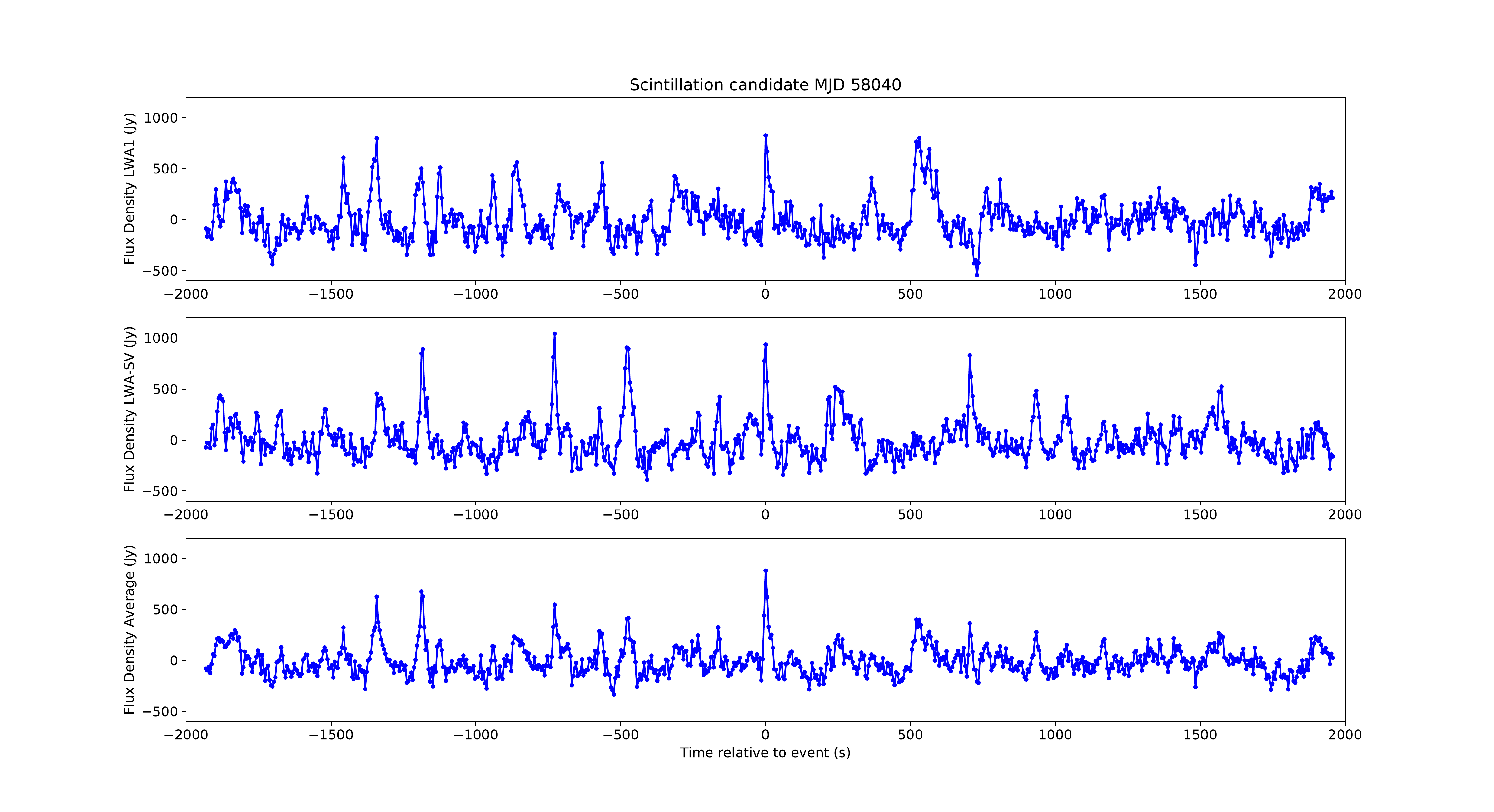}
\caption{Plot showing the light curves of the scintillation candidate MJD 58040 on a longer time axis. The first top panel shows the light curve of event from LWA1 with SNR = 4.25. The middle panel shows the light curve from LWA-SV with SNR = 4.88. The bottom panel shows the average light curve from both stations with a SNR = 6.42. The time zero denotes the peak time of the event }
\end{figure*}

\begin{figure*}

\gridline{\fig{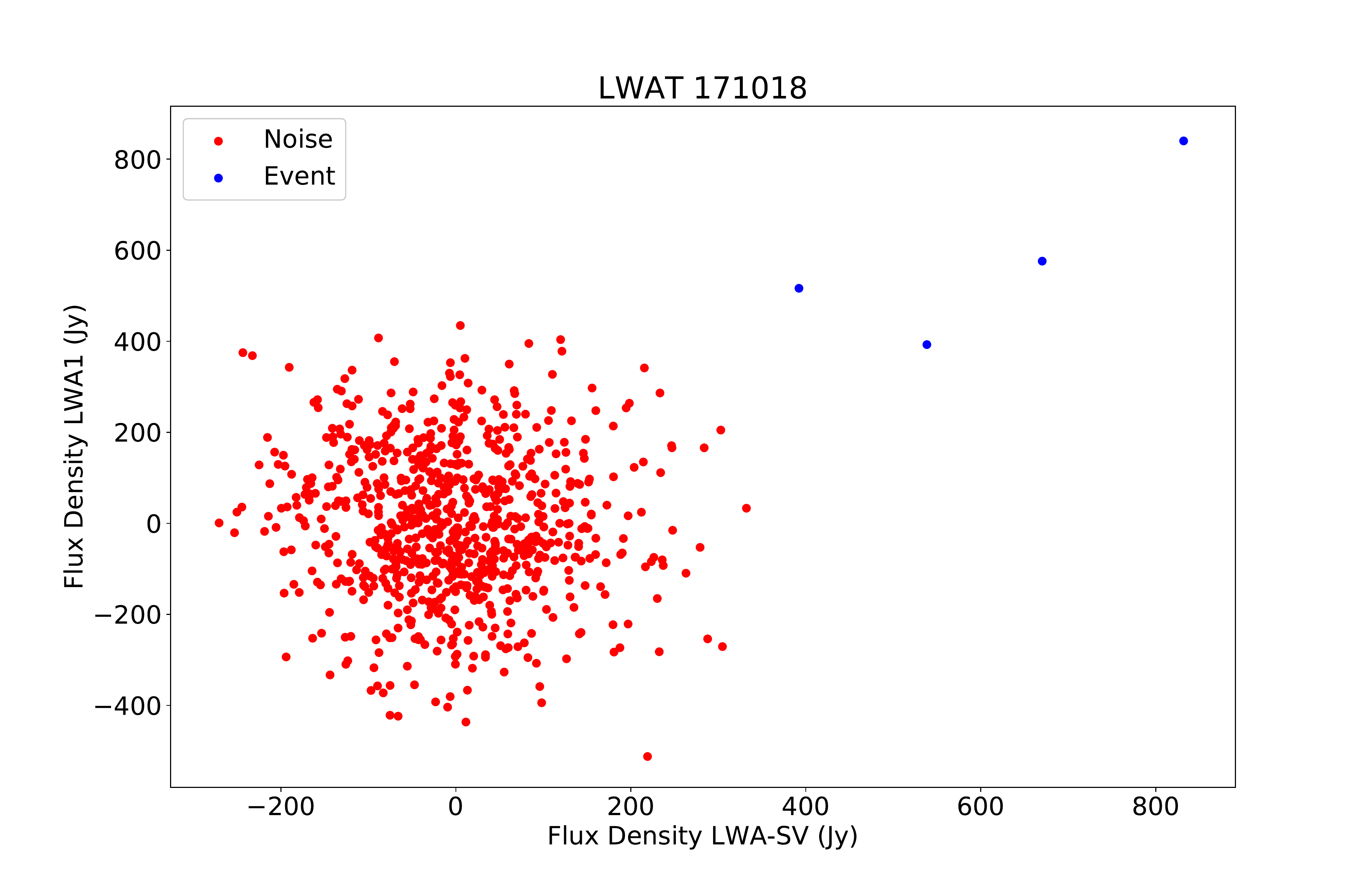}{0.50\textwidth}{}
          \fig{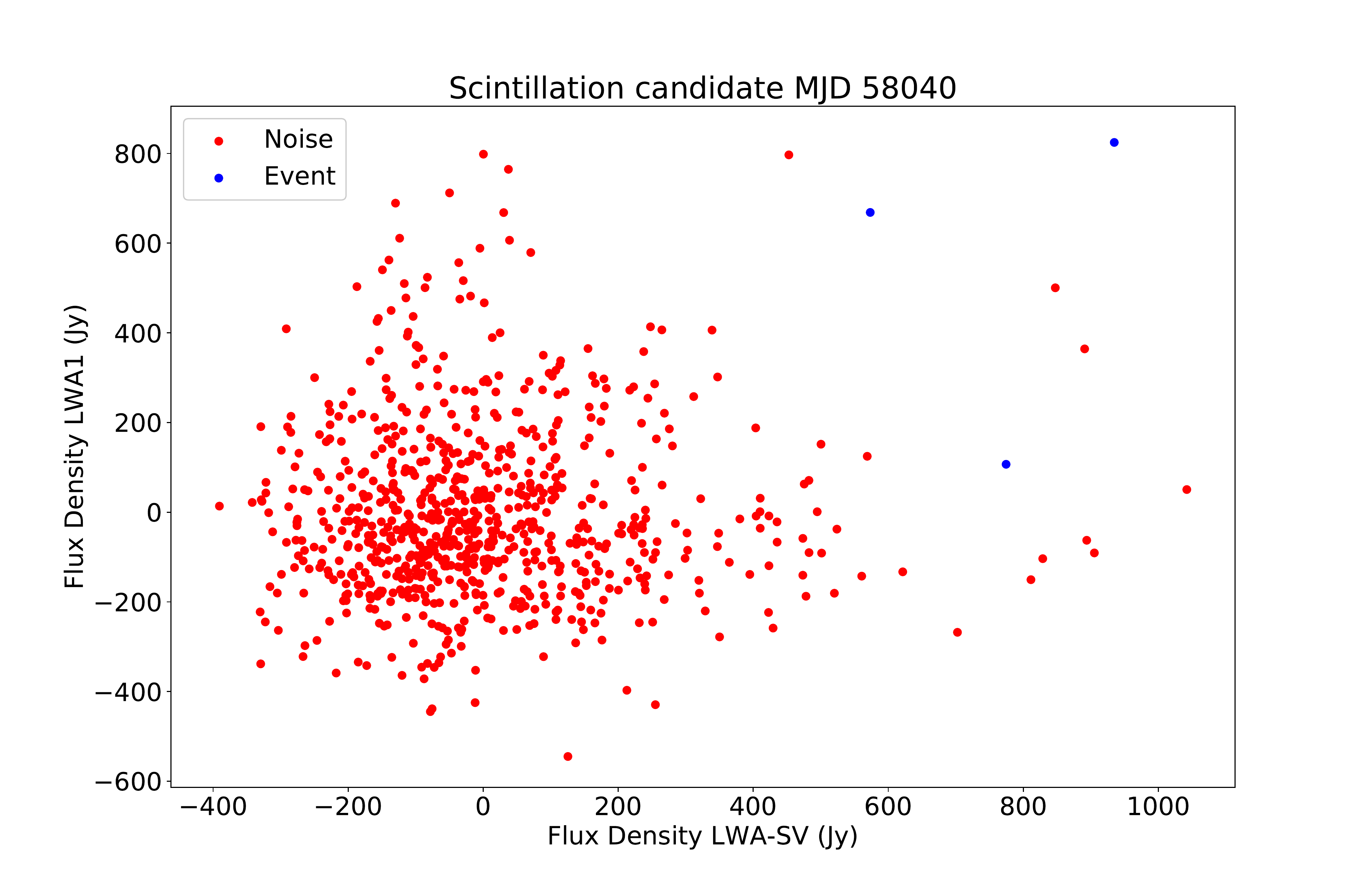}{0.50\textwidth}{}
          }
  
\caption{Scattered plot of the flux density from both stations of LWAT 171018 (left) and   scintillation candidate MJD 58040 (right). The event is defined as the time from 10 s before and after the peak flux point which is denoted as zero second in the light curve. Noise is defined as all the points in light curve other than the event. The red points denote the noise and blue points indicate the transient event.}
\end{figure*}


\begin{figure*}
\gridline{\fig{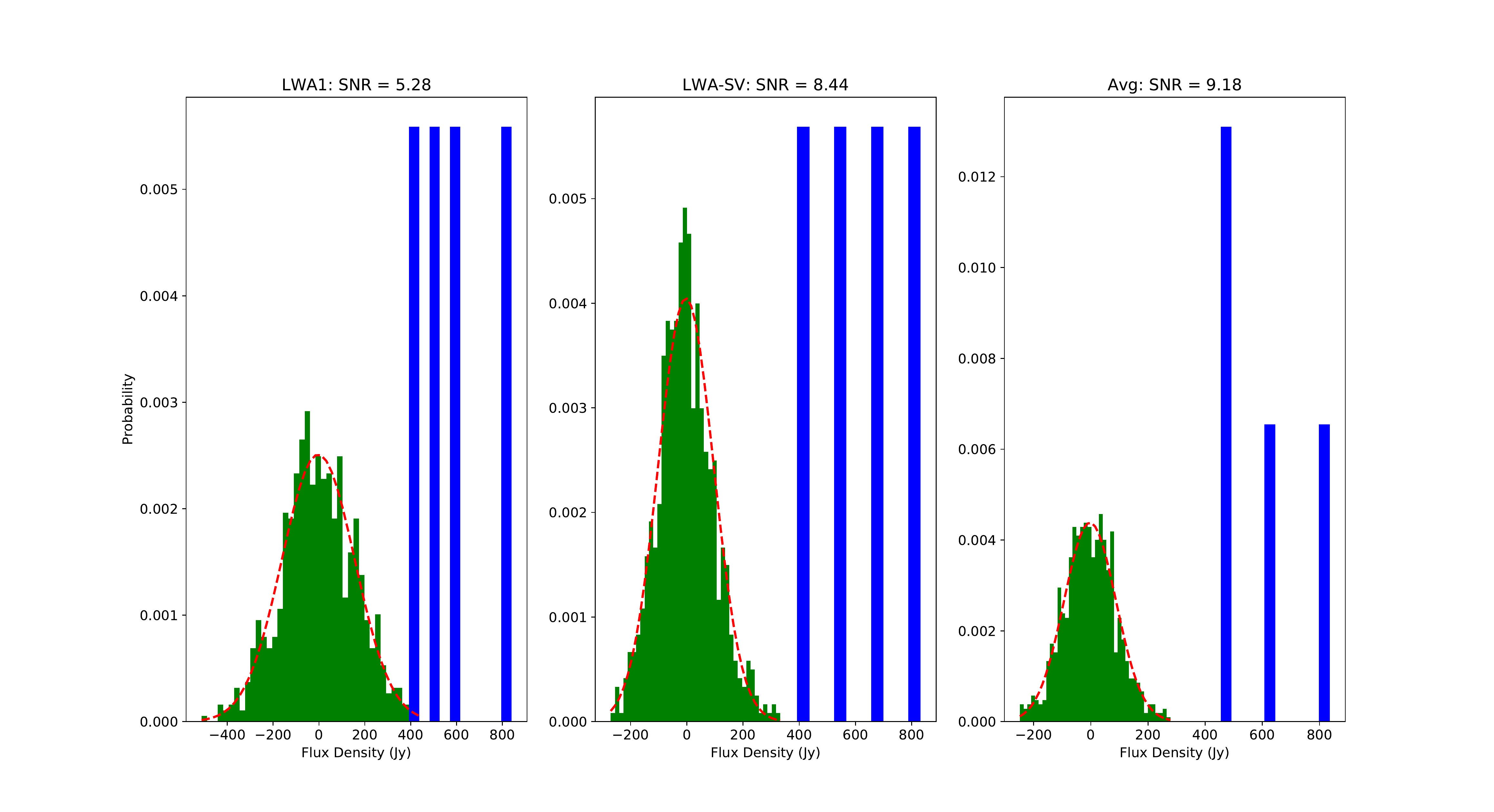}{1.0\textwidth}{}}
\gridline{\fig{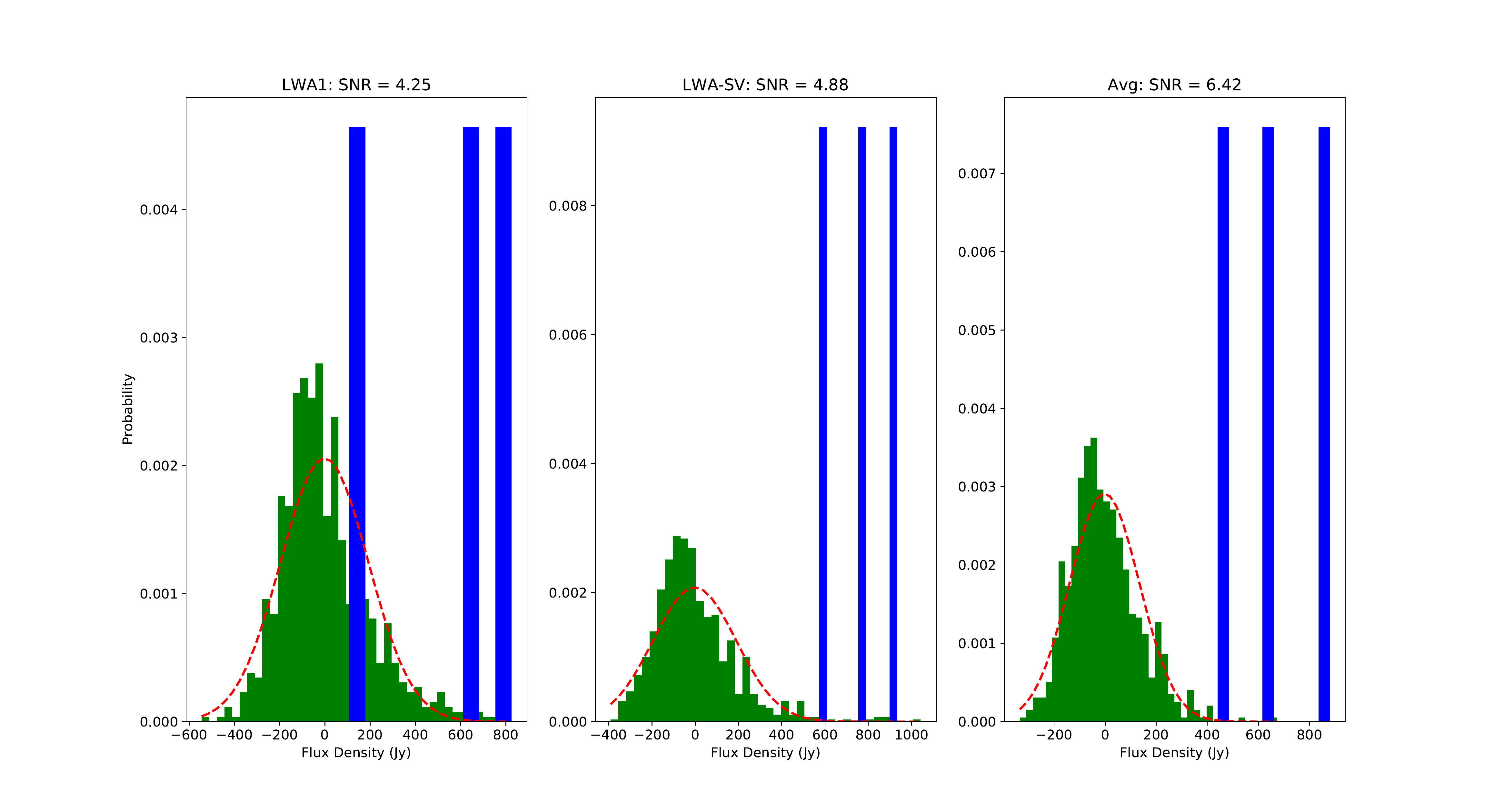}{1.0\textwidth}{}}

\caption{ Histogram plot of the transient event LWAT 171018 (top) and scintillation candidate MJD 58040 (bottom) from the light curves of LWA1, LWA-SV and their average. The green bars show the noise which is fitted with a Gaussian distribution. The blue bars denote the transient event. The calculated SNR ratio is shown on the title of each histogram}
\end{figure*}


Fig. 5 shows the scattered plot of the transient flux density for LWAT 171018 and scintillation candidate MJD 58040 from each station. The plots gives a good estimate of the statistical significance based on the distribution of noise and peak flux for each cases. For LWAT 171018, the noise distribution is clustered and the transient event is well separated from noise suggesting that it is significant. But for scintillation candidate MJD 58040, the noise has a scattered distribution and the transient event is immersed in the noise.

Fig. 6 shows the histogram plots made from the light curves of LWAT 171018 and scintillation candidate MJD 58040 respectively. The histograms fitted with Gaussian profile provide a better picture to understand the distribution of noise and the transient event. The noise is more or less Gaussian in both histograms. The LWAT 171018 is well separated from noise where it is not in scintillation candidate MJD 58040 as the tail of the Gaussian fit goes to higher flux density values.

The analysis of two events based on the light curve pattern, scatter plots, histograms and SNR ratio suggest that the LWAT 171018 is significant and different from the scintillation candidate MJD 58040. Furthermore, it also demonstrates that the LWAT 171018 observed by two stations is not a coincident random spurious signal but a real one. 

In order to characterize the scintillation better, an index or a statistical parameter was necessary. The kurtosis of a probability distribution can be used as an index for measuring the amount of scintillation. In probability and statistics, kurtosis is defined as the ratio of fourth central moment and square of variance. In simple words, kurtosis gives the measure of the infrequent outliers in a distribution. The kurtosis value for a Gaussian distribution in Fisher's definition is zero. Kurtosis of the light curve in each station before and after the event can be calculated to understand how deviant the noise is from a Gaussian distribution. If we use the kurtosis as a measure of scintillation, low kurtosis or close to zero kurtosis events should be scintillation quiet and high kurtosis events should be high scintillation. This exercise was carried out for all the 19 commonly detected events, one hour before and after the peak event and the values are listed in Table 1. Fig. 7 shows the plot of measured kurtosis value for each event in both stations. The LWAT 171018 has a kurtosis value close to zero in both stations whereas all the other events have much higher kurtosis values. There are some scintillation events with high kurtosis value in one station and low kurtosis value in the other station. The high kurtosis value in one station is basically due to the presence of bright, short duration RFI spikes along with the source scintillation. The close to zero kurtosis values in both stations suggests that LWAT 171018 is different from other events and the origin of such a signal is not due to scintillation. 
\begin{figure*}
\epsscale{1.25}
\plotone{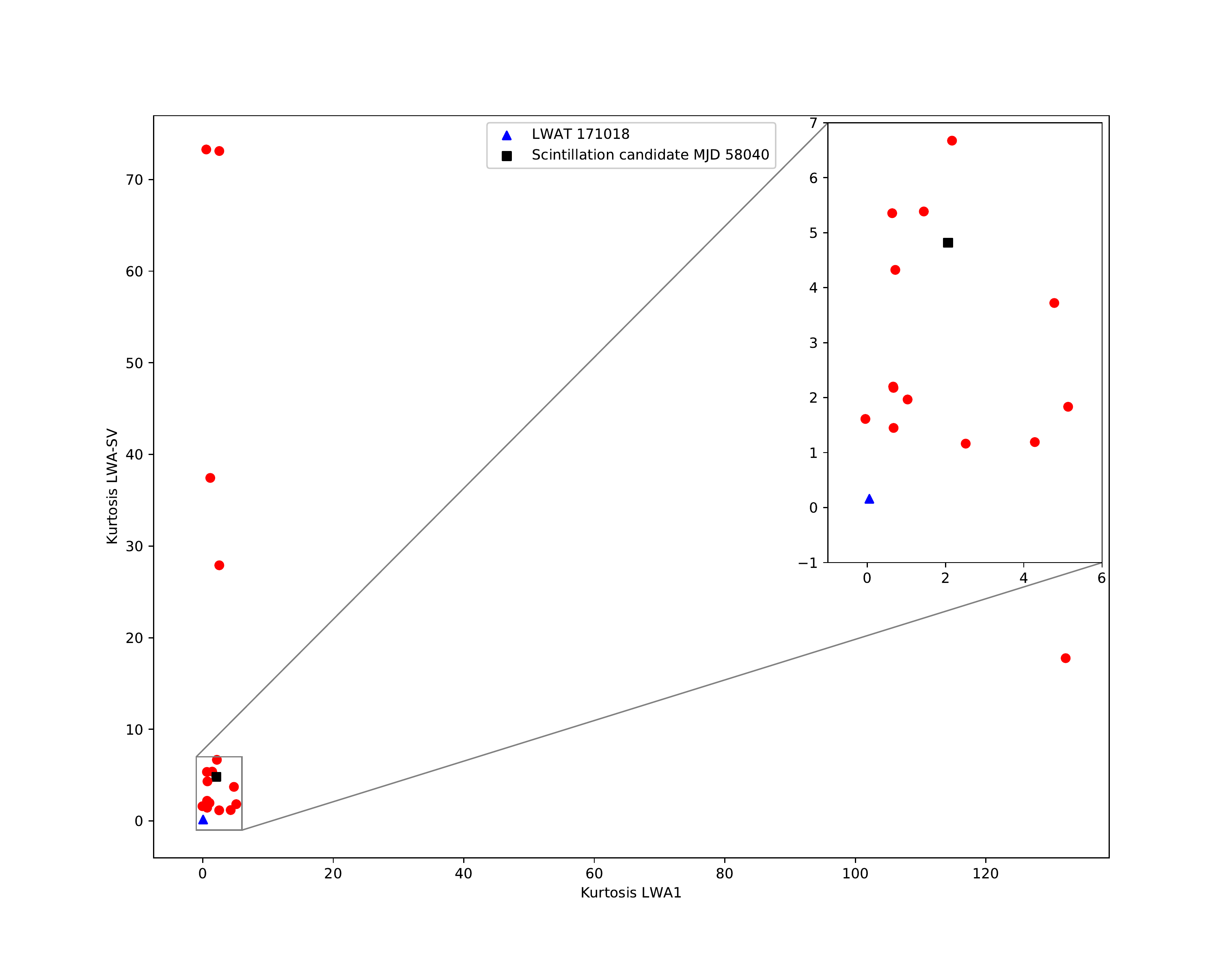}
\caption{ The Kurtosis plot of the light curves from LWA1 and LWA-SV for each common event listed in Table 1. The rectangle shows the zoomed portion of the closely associated points towards origin.}
\end{figure*}

While the source appears statistically separate from scintillating sources, there is a 25 Jy (at 74 MHz) source, 4C +1.06, within error circle plot (see Fig. 9). 4C +1.06 appears to be a 30 arcsec compact radio source from the 20 cm VLA observations \citep{http://adsabs.harvard.edu/abs/1985A26A...148..323R}. The Scintillation of this source has triggered the single station transient pipeline numerous times in the two years of data used in this study. The source has shown up on 15 occasions in LWA-SV and 9 different occasions in LWA1, but LWAT 171018 is the only time a source has shown up in both statons at the same time. Despite the fact that LWAT 171018 appears different from these scintillation events, it remains a possibility that the two coincident events were two, unlucky scintillation spikes from 4C +1.06. We note however, that as mentioned above there is a 15\% chance that a random event will be spatially coincident with a VLSS source bright enough to be detected through a scintillation spike. So the fact that LWAT 171018 is spatially coincident with 4C +1.06 could simply be an unlucky coincidence.  

In order to understand more about the scintillation of 4C +1.06 triggered in each station, the kurtosis one hour before and after the event as well as the peak fluxes were calculated. The 4C +1.06 source was observed to scintillate with an average peak flux of 6.20 $\sigma$ and a kurtosis of 0.66 in LWA1 and with an average peak flux of 6.01 $\sigma$ and a kurtosis of 3.12 in LWA-SV. Fig. 8 shows the histogram of the kurtosis measured during the scintillation of 4C +1.06 occurred on different occasions in each station. The source has experienced low and high scintillation in both stations at different times. But none of the events were measured with a close to zero kurtosis value which was observed for LWAT 171018. This suggests that LWAT 171018 is less likely a  co-incident scintillation spike from 4C +1.06.
\begin{figure*}

\gridline{\fig{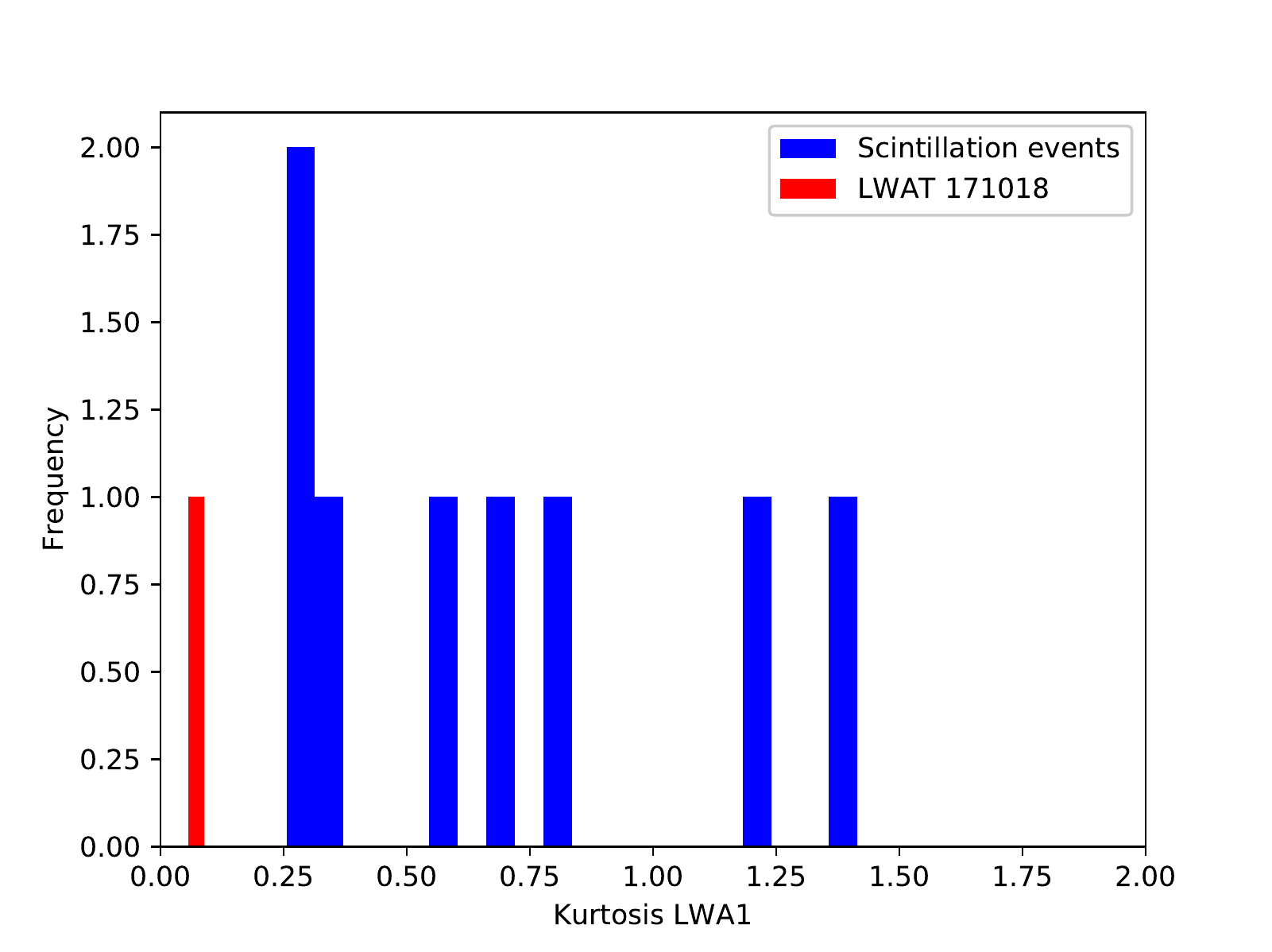}{0.50\textwidth}{}
          \fig{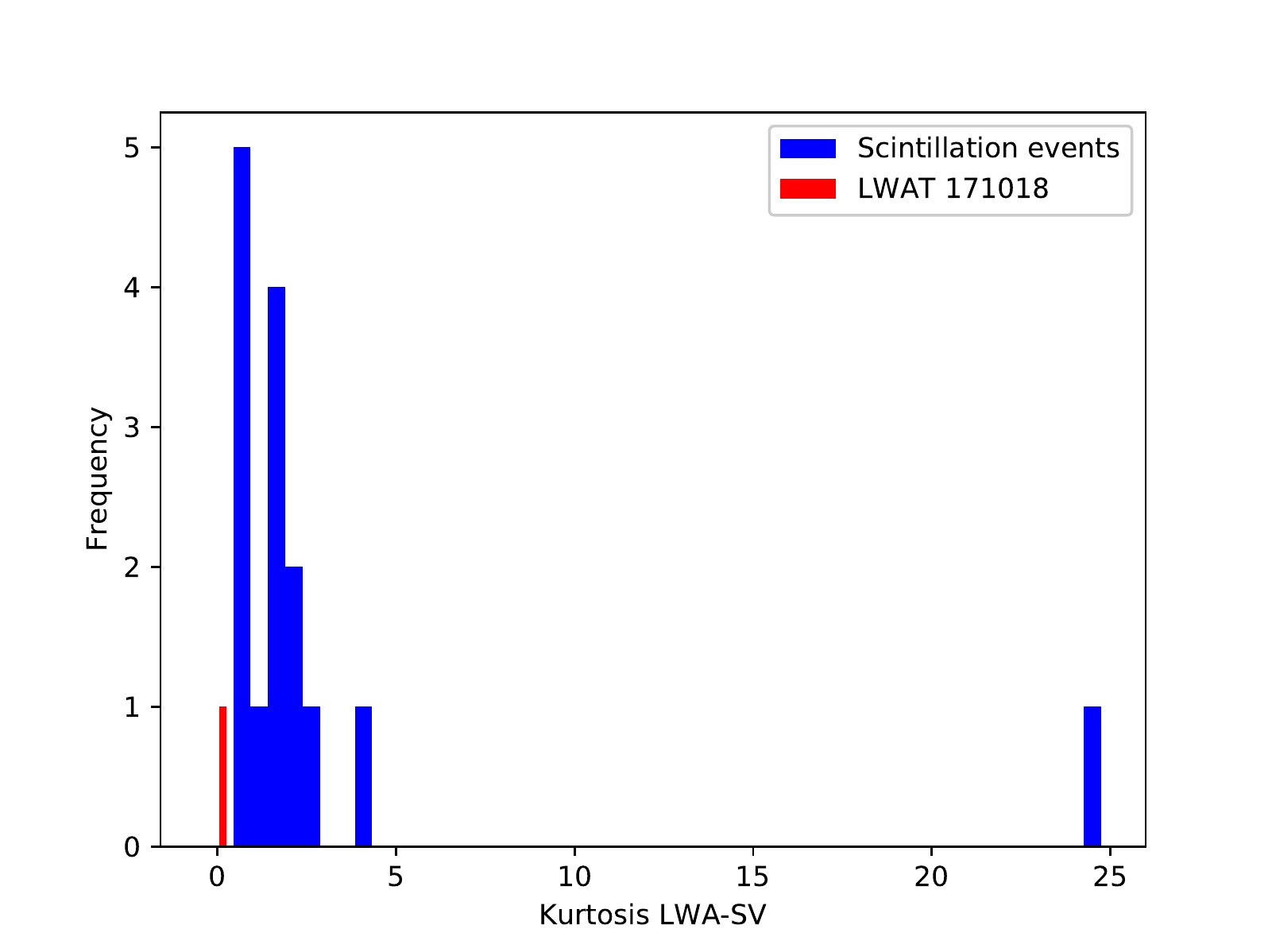}{0.50\textwidth}{}
          }
  
\caption{Histogram plot showing the kurtosis value of the scintillation from 4C +1.06 on different occasions in LWA (left) and LWA-SV (right) over the course of 10,240 h of observation. The kurtosis value of the LWAT 171018 from each station are shown in red.}
\end{figure*}

\paragraph{Satellites}
The next possible candidate is the  reflection or unknown emission from satellites. The low earth orbit satellites can be ruled out as their spatial position changes in the all - sky images. Our transient case is a stationary point source suggesting the possibility of geostationary satellites. Various websites are available on the Internet for tracking the position of satellites. By tracking the position of satellites above the horizon of both stations using In-The-Sky.org website\footnote{\url{https://in-the-sky.org/satmap_worldmap.php\#}}, one candidate satellite was found in the vicinity of the transient. The satellite was Morelos 3, a Mexican communication satellite which is designated to transmit at 1 - 2 GHz and 12 - 18 GHz. 

Reflections from a satellite requires dimensions on the order of a wavelengths. At 34 MHz ($\lambda$ = 9 m), the longest dimension of the fully expanded configuration of the satellite\footnote{\url{http://spaceflight101.com/atlas-v-morelos-3/morelos-3/}} is 41 m (4.5 $\lambda$). While an object of this size is capable of scattering a 34 MHz wave, it is so small that a bright reflection is unlikely. Moreover, the reflection of man-made RFI (the only thing possibly bright enough) would be strongly polarized, which is not the case for the transient reported here.

Alternatively, since transmitters are imperfect, there could be a possible unpolarized out of band emission from satellite transmitters at lower frequencies. As of now, we do not know the origin of any such emission mechanisms. The Morelos 3 was launched in October 2015 and both LWA stations have been collecting all - sky images since May 2016. If this was a signal from the satellite, one or both stations would likely see the signal at other times. In order to check for any kind of previous signals from geostationary satellite, the all-sky image from both stations were searched at the corresponding azimuth and altitude locations. We could not find a single case of emission at the position of the satellite.

Fig. 9 shows the 1-$\sigma$ position error circle plot from each station along with the location of transients, satellites, VLSS sources 4C +1.06, NGC 1218, 4C +04.11 and an optical supernova detected in the vicinity.
\begin{figure}
\epsscale{1.25}
\plotone{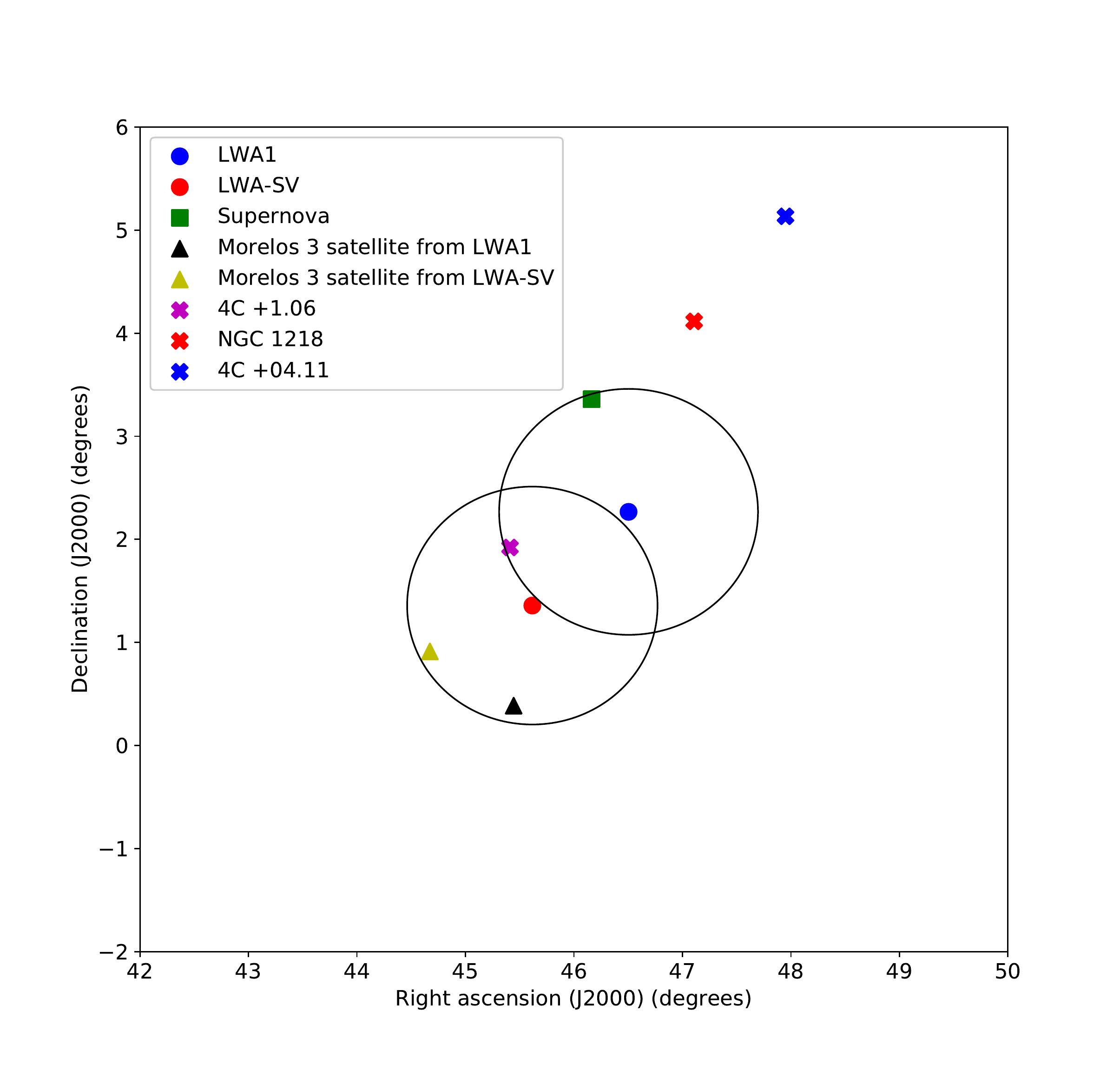}
\caption{Plot showing 1-$\sigma$ position error circle centered on the transient location from each station. The location of satellites seen from each station, optical supernova and the VLSS sources 4C +1.06, NGC 1218 and 4C +04.11 is shown. The geometry is considered to be flat as the area shown is only $8^{\circ} \times 8^{\circ} $ in extent. }
\end{figure}
The position error for each telescope takes into account of the pointing error of the telescope, signal to noise error and the random error due to ionospheric fluctuation at low frequencies. This estimated value of position error was 1.19 degrees for LWA1 and 1.15 degrees for LWA-SV. 

\section{Discussion}

\subsection{Optical or high-energy counterparts}

Having ruled out all the known cases of false positive events, we are left with a previously undiscovered cosmic signal. We searched for any optical or high-energy counterparts, including gamma ray bursts, flare stars, bright radio sources and standard supernovae. We noticed that a standard supernova went off in the same direction (03:04:39.35, +03:21:32.52) of the sky on the same day at 11:38:24 UTC. The optical supernova, AT 2017hps was detected by the Pan-STARRS1 group (Transient Name Server\footnote{\url{https://wis-tns.weizmann.ac.il/object/2017hps/discovery-cert}}; The Open Supernova Catalog\footnote{ \url{https://sne.space/}}; \citealt{2017ApJ...835...64G}) and the location of the transient is marked on the position error plot (See Fig. 9). A standard supernova occurs frequently in all directions of the sky and the possibility of low frequency radio emission from them is not clear. 46 supernovae were detected within $\pm$2 days of the event in different parts of the sky with a declination greater than $-25$ degrees. 

An estimate of the probability can be calculated by assuming that all the supernovae events occurred  randomly in the sky 30 degrees above the horizon. For this purpose, the radius of the error circle is the position error in LWA1 which is 1.19 degrees (1/48 rad).

\begin{equation}
Probability = \left[\frac{\mbox{Number of  supernovae events}}{\mbox{Number of beams (error circles)}}\right]
\end{equation}
\begin{eqnarray}
Number\: of\: beams = \left[\frac{\mbox{Area of LWA sky}}{\mbox{Area of beam}}\right] \\
= \frac{\pi \mbox{ rad}^{2}}{\pi \left(\frac{1}{48} \mbox{ rad}\right)^{2}} = 2304
\end{eqnarray}
\begin{equation}
Probability = \frac{46}{2304} = 1.98\; \%
\end{equation}

So the probability of a standard supernova to occur within the positional error circle of both LWA station is 1.98\%.

\subsection{New radio transient}
The lack of evidence supporting the false positive detections and the absence of any clear optical or high-energy counterpart suggest that this could possibly be a new type of cosmic event.  Previous observations have not detected any such kind of high flux density transient events at low radio frequencies. We have only detected this one event since LASI began producing all-sky images from both stations in May 2016. With respect to the data used for this study, the new transient source does not repeat. The nature of the transient is not clear as it lacks other EM counterparts and has only occurred once.

The single cosmic transient event was detected after 10,240 hours of observation. Each hour has 720 all -sky images and each image has 2304 independent beams in the sky 30 degrees above the horizon. This makes a total of $1.70 \times 10^{10} $ independent beams. For Gaussian statistics, the probability of finding a 5.28 $\sigma$ detection in LWA1 is $6.46 \times 10^{-8}$ and 8.46 $\sigma$ in LWA-SV is $1.34 \times 10^{-17}$. The joint probability of finding such an event simultaneously in both stations is given by their product which is $8.64 \times 10^{-25}$.  The expected number of such events we should have seen is given by the product of the joint probability and number of independent beams.  For the $1.70 \times 10^{10} $ independent beam integrations observed with both stations the calculated number of such events is 
 $\approx 1.47 \times 10^{-14}$.  This is much less than one implying that this is a real event and not just a chance occurrence of two simultaneous noise peaks.

The radio waves traveling through the ionized plasma in the intergalactic medium cause a difference in the arrival time of signals. Higher frequency signals will arrive first and the measured pulse over a frequency bandwidth will be dispersed. An upper limit of the dispersion measure (integrated electron density along the line of sight) can be calculated using the pulse width of the transient from light curve. The dispersion measure is calculated on the assumption that the pulse is dispersed along 100 kHz bandwidth of the TBN data.  The relation between time delay in the arrival of two different frequencies and dispersion measure is given by 

\begin{equation}
\Delta t_{d} = 4.149 \times 10^{3}\: \frac{ \rm MHz^{2}\; cm^{3}\; s}{\rm pc}  \left(\frac{1}{f_{1}^{2}} - \frac{1}{f_{2}^{2}}\right)\; DM
\end{equation}

In the above equation, $DM$ is the dispersion measure, $\Delta t_{d}$ was taken to be 15 s from light curve, $f_{1}$ = 33.95 MHz and $f_{2}$ = 34.0375 MHz. Putting all these values in equation (5) will return a $DM$ = 804 pc cm$^{-3}$. If we compare the $DM$ value with the known transient sources, it will fall into the group of recently detected Fast Radio Burst (FRB) events.
\citet{http://adsabs.harvard.edu/abs/2007Sci...318..777L}  detected the first FRB in 2007 at 1.4 GHz after analyzing the archival survey data of Magellanic clouds using Parkes Radio telescope in Australia. The burst had a flux density of 30 Jy and a duration 5 ms. The pulse was dispersed with a DM of 375 pc cm$^{-3}$ and was far away from the Galactic plane suggesting an extra galactic origin.  In later years, further observations using Parkes, GBT, Arecibo have discovered over 17 FRBs at high frequencies and these are listed in the FRB catalog \citep{2016PASA...33...45P}. Recently the CHIME/ FRB Project has discovered 13 new FRBs at frequencies between 400 and 800 MHz in their pre-commissioning phase. One of the detetced FRBs was observed to have 6 repeated bursts. The hypothesized origin of these short bursts was thought to be exotic phenomena like merging neutron stars or evaporating blackholes. The detection of repeating bursts eliminates the cataclysmic models for the FRB source \citep{https:// doi.org/10.1038/s41586-018-0867-7, https://doi.org/10.1038/s41586- 018-0864-x}.  However, no FRBs have been detected below 100 MHz.

An FRB is potentially a good candidate for LWAT 171018. Pulse broadening can occur at lower frequencies due to dispersion and scattering causing seconds of time delay. Since the calculated upper limit of DM is high and the source location is far away from the Galactic plane $(l = 176.13, b = -46.88)$, the transient could be possibly an extra galactic source. 

The expected scattering width at 34 MHz can be calculated using the relation
\begin{equation}
\tau_{sc}(\nu) \propto \nu^{\gamma}
\end{equation}
where $\tau_{sc}$ is the scattering time scale and $\gamma$ is the scattering index which is taken to be --4 for this case. For a short duration, $<$ 1.1 ms FRB pulse from \citet{http://adsabs.harvard.edu/abs/2013Sci...341...53T} at 1.3 GHz, the estimated pulse width at 34 MHz is $\approx$ 2400 s. The measured 15 s pulse width from light curve is much less than the expected pulse width due to scatter broadening. 

Several other imaging campaigns have been conducted at low frequencies for FRB detection. \citet{http://dx.doi.org/10.1088/0004-6256/150/6/199} searched for FRBs using MWA between 139 -170 MHz. No FRBs were detetced in the 2 s de-dispersed images collected over 10.5 hours of observation covering 400 square degrees. This search placed a limit of $ < $ 700 events/day/sky  within the flux density limit of 700 Jy for a DM of 170-675 $\rm pc/cm^{3}$. \citet{10.1093/mnras/stw451} conducted a survey for transient searches at 182 MHz with MWA using 28 s integration images. No FRBs were detected within the flux density limit of 0.285 Jy. The survey placed  an upper limit of $<$ 82 FRBs/day/sky within the flux density limit of 7980 Jy for a DM $<$ 700  $\rm pc/cm^{3}$. \citet{https://doi.org/10.3847/2041-8213/aae58d} conducted a coordinated MWA observations to shadow the low frequency component of the FRBs detected by the Australian Square Kilometre Array Pathfinder (ASKAP) at 1.4 GHz. The simultaneous MWA observations of 7 ASKAP FRBs between 70-200 MHz using 0.5 s images did not detect any low frequency emission. The results from previous observations and smaller pulse width compared to the expected width from scattering implies that the observed transient is less likely to be an FRB event. 

This also implies that the detected transient is new and at the same time it is similar to the transient detected by  \citet{2016MNRAS.456.2321S}, ILT J225347+862146. The 6 sigma detection threshold of 38 MHz LASI images at zenith is 250 Jy \citep{2015JAI.....450004O}. The sensitivity of an 11 minute LOFAR MSSS image is greater than 7.9 Jy. The detected ILT J225347+862146 had a flux density of 20 Jy and 11 minute duration. The LASI images are not sensitive enough to detect ILT J225347+862146. At the same time, LOFAR MSSS images could have easily detected LWAT 171018 as it was a 800 Jy bright event. But if  ILT J225347+862146 lasted only for 5 s duration and assuming the fluence is same at 60 and 38 MHz, then the peak flux density of  the event would be 2640 Jy. This event could be easily observed in LASI images. In the same way, if the 800 Jy LWAT 171018 lasted for 11 minute in MSSS images, then the peak flux would be 19.09 Jy which is also above the detection threshold.

\subsection{Burst Location}
An upper bound on the distance using the DM can be written as $DM \approx 1200 \;z$ pc cm$^{-3}$ \citep{http://adsabs.harvard.edu/abs/2007Sci...318..777L}. So a DM of 804 pc cm$^{-3}$ can give  a redshift, $z \approx  0.67$ 
The observed contribution of $DM$ from Milky Way is less than 100 pc cm$^{-3}$ for Galactic latitudes greater than 10 degree \citep{http://dx.doi.org/10.3847/2041-8205/830/2/L31}. The total observed $DM$ is the sum of the contribution from the host galaxy intergalactic medium and that of the Milky Way \citep{ju}. After removing the contribution from the Milky Way a $DM$ of 700 pc cm$^{-3}$ gives redshift, $z \approx 0.58$. This is an upper limit of the redshift solely based on the temporal pulse width of the transient event. 

\section{Conclusions}

By using two LWA stations separated by 75 km we present an anti-coincidence study of the joint observations over a period of 10,240 hours between May 2016 and July 2018.  During this period nineteen events were detected simultaneously from both stations in the same part of the sky, however all but one of these can be classified as the result of scintillation of a known compact radio source induced by the ionosphere or RFI. One source on 18 October 2017 with a flux density of 840 Jy at 34 MHz is not readily explained by scintillation or RFI.  After ruling out a number of possible origins we find that this new transient could be a previously unknown cosmic signal. The origin of this source is not clear due to the lack of evidence. 

Multi epoch observations using sensitive telescopes at low frequencies may yield further emission signals if the transient source is still active. In the future, we will continue the all-sky monitoring  to search for similar cosmic transient events using both LWA stations. Multi wavelength observations of cosmic transient sources followed by an LWA trigger could provide insights into the source structure and process of emission mechanisms.  Future observations of similar transients will also benefit from the implementation of a broadband (10 MHz) all sky correlator that now runs continuously at the LWA-SV station.  

\acknowledgments
Construction of the LWA has been supported by the Office of Naval Research under Contract N00014-07-C-0147 and by the AFOSR. Support for operations and continuing development of the LWA1 is provided by the Air Force Research Laboratory and the National Science Foundation under grants AST-1711164 and AGS-1708855.

%

\vspace{5mm}
\facilities{LWA1, LWA-SV, Pan-STARRS}


\software{LWA Software Library \citep{2012JAI.....150006D}}

\end{document}